\documentclass[aps,pra,reprint,groupedaddress]{revtex4-1}

\usepackage[english]{babel}
\usepackage{multirow}
\usepackage{amsmath, amstext, amsfonts, amssymb}
\usepackage{graphicx}
\usepackage{hhline}
\usepackage{colortbl}
\usepackage[dvipsnames]{xcolor}
\usepackage[normalem]{ulem}
\usepackage{physics}
\usepackage{subdepth}
\usepackage{color}
\usepackage[nolist]{acronym}
\definecolor{Gray}{gray}{0.85}

\definecolor{LightCyan}{rgb}{0.88,1,1}

\newcommand{\sep}{\cdot}
\newcommand{\supp}{Supplement:}
\newcommand{\bt}[1]{\textcolor{black}{#1}}

\begin{document}

\preprint{commit-1}
%
%
\title{Multi-reference quantum chemistry protocol for simulating autoionization spectra: Test of ionization continuum models for the neon atom}

%
%
%
%
\author{Gilbert Grell}
\affiliation{Institut f\"ur Physik, Universit\"at Rostock, Albert-Einstein-Str. 23-24, 18059 Rostock, Germany}
%
%
\author{Oliver K\"uhn}
\affiliation{Institut f\"ur Physik, Universit\"at Rostock, Albert-Einstein-Str. 23-24, 18059 Rostock, Germany}
%
%
%
\author{Sergey I. Bokarev}
\email[]{sergey.bokarev@uni-rostock.de}
\affiliation{Institut f\"ur Physik, Universit\"at Rostock, Albert-Einstein-Str. 23-24, 18059 Rostock, Germany}
\date{\today}
%
%
\begin{abstract}
  In this contribution we present a protocol to evaluate partial and total Auger decay rates combining the restricted active space self-consistent field electronic structure method for the bound part of the spectrum and numerically obtained continuum orbitals in the single-channel scattering theory framework.
  On top of that, the two-step picture is employed to evaluate the partial rates.
  The performance of the method is exemplified for the prototypical Auger decay of the neon $1s^{-1}3p$ resonance.
  Different approximations to obtain the continuum orbitals, the partial rate matrix elements, and the electronic structure of the bound part are tested against theoretical and experimental reference data.
  It is demonstrated that the partial and total rates are most sensitive to the accuracy of the continuum orbitals.
  For instance, it is necessary to account for the direct Coulomb potential of the ion for the determination of the continuum wave functions.
  The Auger energies can be reproduced quite well already with a rather small active space.
  Finally, perspectives of the application of the proposed protocol to molecular systems are discussed.
\end{abstract}
%
\pacs{}
%
\maketitle
%
%
\begin{acronym}[RASSCF]
 \acro{SCF}{Self-Consistent Field}
 \acro{PT2}{Second Order Perturbation Theory}
 \acro{RAS}{Restricted Active Space}
 \acro{CAS}{Complete Active Space}
 \acro{AS}{Active Space}
 \acro{MCSCF}{Multi-configurational \ac{SCF}}
 \acro{RASSCF}{\ac{RAS} \ac{SCF}}
 \acro{CASSCF}{\ac{CAS} \ac{SCF}}
 \acro{RASPT2}{Restricted Active Space Second-order Perturbation Theory}
 \acro{RASSI}{\ac{RAS} State Interaction}
 \acro{HF}{Hartree-Fock}
 \acro{CI}{Configuration Interaction}
 \acro{MCDF}{Multi-configurational Dirac-Fock}
 \acro{ADC}{Algebraic Diagrammatic Construction}
 \acro{FANO-ADC}[Fano-ADC]{Fano-Stieltjes Algebraic Diagrammatic Construction}
 \acro{DKH}{Douglas-Kroll-Hess}
 \acro{X2C}{Exact Two Component Decoupling}
 \acro{ANO}{Atomic Natural Orbital}
 \acro{QC}{Quantum Chemistry}
 \acro{GTO}{Gaussian Type Orbital}
 \acro{ICD}{Interatomic Coulombic Decay}
 \acro{ETMD}{Electron Transfer Mediated Decay}
 \acro{PES}{Photoelectron Spectroscopy}
 \acro{AES}{Auger Electron Spectrum}
 \acro{AIS}{Autoionization Spectroscopy}
 \acro{RPES}{Resonant \ac{PES}}
 \acro{SE}{Schr\"odinger Equation}
 \acro{SD}{Slater Determinant}
 \acro{SO}{Strong Orthogonality}
 \acro{GTO}{Gaussian Type Orbital}
 \acro{DO}{Dyson Orbital}
 \acro{MO}{Molecular orbital}
 \acro{GS}{Gram-Schmidt}
\end{acronym}

%
%
%
\section{Introduction}
\label{sec_intro}
\begin{figure}[b]
  \includegraphics{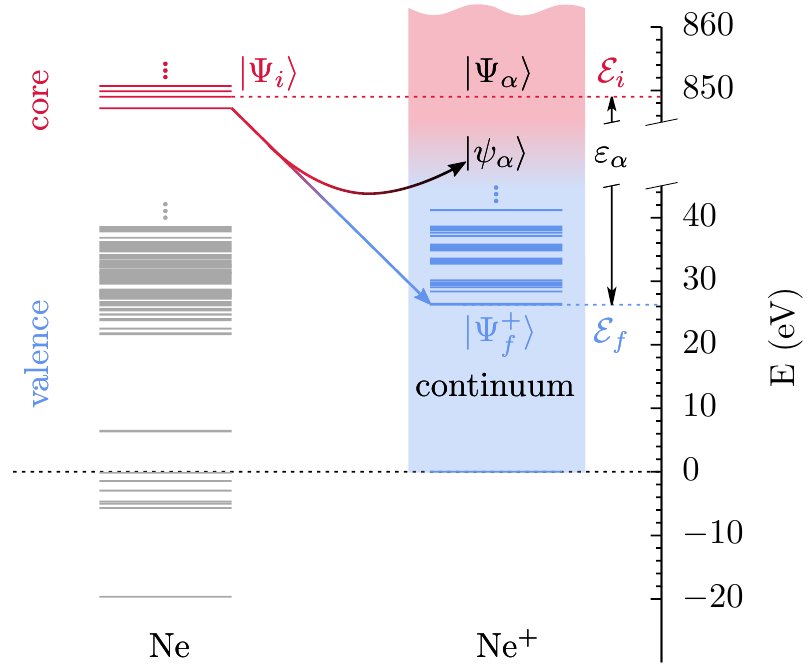}
  \caption{\label{fig:intro} Autoionization scheme for the neon atom. The core vacancy state $\ket{\Psi_i}$ with energy $\mathcal{E}_i$ (red) decays isoenergetically into the continuum state $\ket{\Psi_\alpha}$ (black) composed of the ionic bound state $\ket{\Psi_f^+}$ with energy $\mathcal{E}_f$ (blue) and the continuum orbital $\ket{\psi_\alpha}$ of the outgoing electron with the excess energy $\varepsilon_\alpha$.
  States that do not contribute to the process are depicted in gray; the singly ionized continuum is denoted by the color gradient.
          }
\end{figure}

%
%
Ionization triggered by photon absorption occurs along two pathways.
In direct photoionization, the energy is transferred to an ejected electron.
Alternatively, the system can be first put into a metastable state by a resonant excitation and afterwards decay via an autoionization mechanism.
%
%
Autoionization can be approximately understood as a two-step process \cite{Wentzel_ZP_1927}, in which the decay can be considered independently from the excitation process and interferences between direct and autoionization are neglected.
For example, let us consider an atomic species, such as a neon atom that is prepared in a highly excited state $\ket{\Psi_i}$ above the continuum threshold at $E=0$~eV, Fig.~\ref{fig:intro}.
This state spontaneously decays into the continuum state $\ket{\Psi_\alpha}$ comprising the discrete state $\ket{\Psi_f^+}$ of the ion and the emitted electron, $\ket{\psi_\alpha}$, carrying the excess energy $\varepsilon_{\alpha}=\mathcal{E}_i - \mathcal{E}_f$.
%
%
The system's electronic structure is thus encoded into the kinetic energy spectrum of the ionized electrons.
%
\ac{PES} and \ac{AIS} map bound states to the continuum which makes them less sensitive to selection rule suppression and more informative than spectroscopies involving optical transitions between bound states~\cite{VanderHeide2012, Hofmann2013, Huefner_Book}.
%

%
%
Autoionization processes, predominantly Auger decay~\cite{Meitner_ZP_1922}, but also \ac{ICD}~\cite{Cederbaum_PRL_1997} and \ac{ETMD}~\cite{Zobeley_JCP_2001} are particularly interesting on their own.
Due to their correlated nature, they not only probe but also initiate or compete with intricate ultrafast electronic and nuclear dynamics e.g.~\cite{Slavicek_JPCL_2016, Slavicek_JACS_2014, Brown_ARSCPC_2009, Stumpf_NC_2016, Wang_PRA_2018}.
Additionally, they provide the main channel for the decay of core vacancies~\cite{deGroot_Book}
and play a key role in biological radiation damage, creating highly charged cations while a cascade of highly reactive low energy electrons is emitted~\cite{Howell_IJRB_2008, Alizadeh_ARPC_2015, Stumpf_NC_2016, Martin_IJRB_2016, Yokoya_IJRB_2017}.
Further, in free electron laser experiments operating with ultrashort intense X-ray pulses, autoionization after multiple photoionization induces the Coulomb explosion of the target, which limits the achievable spectroscopic and temporal resolution~\cite{Neutze_N_2000}.
%
%
Due to this wealth of applications, autoionization and especially the local Auger effect have been studied extensively both theoretically and experimentally since its discovery by Meitner~\cite{Meitner_ZP_1922} and description by Wentzel~\cite{Wentzel_ZP_1927}.
Remarkably, \ac{AIS} simulations of molecular systems remain challenging until today, although the fundamental theory is known for decades~\cite{Aberg_CaRiMI_1982, Aberg_PS_1980, Fano_PR_1961b}.
For atoms, methods combining highly accurate four-component \ac{MCDF} calculations with multichannel scattering theory are publicly available~\cite{Fritzsche_CPC_2012}, whereas no such general purpose code exists for molecules.
The main complication of the molecular case lies in the construction of molecular continuum states $\ket{\Psi_\alpha}$.
%
%
The approaches to the simulation of \ac{AIS} published during the last decades can be classified into two families -- those that circumvent the continuum orbital problem and those that treat the continuum orbital explicitly.
%

%
%
The first family comprises the following flavors:
The simplest method that allows to assign experimental \ac{AIS} is to evaluate the energetic peak positions~\cite{Agren_CPL_1975a, Agren_JCP_1981, Hillier_MP_1976}.
On top of that simple estimates for the partial decay rates can be obtained based on an electron population analysis~\cite{Mitani_JESRP_2003}.
More advanced approaches rely on an implicit continuum representation with Stieltjes imaging~\cite{Carravetta_JCP_2000a}, a Green's operator \cite{Schimmelpfennig_JPBAMOP_1992, Schimmelpfennig_JESRP_1995},
or a propagator~\cite{Oddershede_AiCP_1987, Liegener_CPL_1982} formalism.
%
%
From this group, the \ac{FANO-ADC} method~\cite{Averbukh_JCP_2005a} has been used to evaluate
Auger, \ac{ICD} and \ac{ETMD} decay rates of van der Waals clusters~\cite{Kolorenc_JCP_2008}, first row hydrides~\cite{Kolorenc_JCP_2011} and the $[\text{Mg}(\text{H}_2\text{O})_6]^{2+}$ cluster~\cite{Stumpf_NC_2016}.
%
%
\bt{
Therein, the continuum is approximately represented with spatially confined basis functions.
However, the description of Auger electrons with kinetic energies of several hundreds of eV requires large basis sets leading to computationally demanding simulations.
}
The second family, relying on an explicit representation of the continuum wave function, consists of the following approaches:
The one-center approximation which uses atomic continuum functions centered at the vacancy-bearing atom to describe the outgoing electron in the evaluation of partial decay rates~\cite{Siegbahn_CPL_1975b, Jennison_CPL_1980, Larkins_AJP_1990, Fink_JESRP_1995a}.
This approximation can be applied on top of high-level electronic structure methods~\cite{Travnikova_CPL_2009a}. 
%
%
%
%
%
\bt{
It is also applied in the XMOLECULE package~\cite{Inhester_PRA_2016, Hao_SD_2015} which is based on very cost efficient electronic structure calculations and atomic continuum functions.
This allows for the evaluation of ionization cascades but may limit the applicability for strongly correlated systems, e.g., possessing a multi-configurational character.
}
Further, the influence of the molecular field may be taken into account perturbatively~\cite{Faegri_PRA_1979, Higashi_CP_1982} or in a complete manner with, e.g., the single-center approach, where the whole molecular problem is projected onto a single-centered basis
\bt{\cite{Zahringer_PRA_1992, Demekhin_OS_2007, Demekhin_PRA_2009, Demekhin_JCP_2011, Inhester_JCP_2012, Banks_PCCP_2017}.}
Finally, multi-channel scattering theory methods that combine finite multi-centered basis sets with the appropriate boundary conditions to represent the molecular continuum have been developed~\cite{Colle_PRA_1989} and applied to the \ac{AIS} of a variety of small systems~\cite{Colle_PRA_1989, Colle_PRA_1993, Colle_JPBAMOP_2004}.
%
%
For instance, the recently developed XCHEM approach~\cite{Marante_JCTC_2017} has been applied to simulate \ac{PES} and \ac{AIS} of atoms~\cite{Marante_PRA_2017} and small molecular systems~\cite{Klinker_PRA_2018}.
These techniques represent the most general and accurate quantum-mechanical treatment of the problem, thus potentially serving as a high-level reference, although connected to substantial computational effort.
Summarizing, most of the mentioned methods have been applied only to simple diatomics, first row hydrides, halogen hydrides, and small molecules consisting of not more than two heavy atoms.
%
%
%
%
%
%
%
Studies of larger molecular systems, such as tetrahedral molecules, small aldehydes, and amides~\cite{Ortiz_JCP_1984, Correia_JCP_1991}, solvated metal ions~\cite{Stumpf_NC_2016} and polymers~\cite{Endo_JMS_2016} are very scarce.
In fact, the \ac{FANO-ADC}~\cite{Averbukh_JCP_2005a, Kolorenc_JCP_2008} and the XMOLECULE \cite{Hao_SD_2015, Inhester_PRA_2016} approaches are the only publicly available tools that allow to simulate \ac{AIS} for a variety of systems without restricting the molecular geometry.
Further, both methods are not suited to treat systems possessing multi-configurational wave functions.
This puts studies of some chemically interesting systems having near-degeneracies, for example, transition metal compounds, or of photodynamics in the excited electronic states, e.g., near conical intersections, out of reach.
%
%
%
%
To keep up with the experimental advancements, the development of a general purpose framework to evaluate autoionization decay rates (Auger, \ac{ICD}, and \ac{ETMD}) for molecular systems is warrant.
Such a framework should be kept accessible, transferable and easy to use, i.e. it should be based on widespread robust and versatile \ac{QC} methods.
Here, we present a protocol that combines multi-configurational \ac{RASSCF} bound state wave functions with single-centered numerical continuum orbitals in the single-channel scattering theory framework \cite{Burke2011a}.
We have chosen the \ac{RASSCF} approach, since it is known to yield reliable results for core-excited states~\cite{Bokarev_WIRCMS_2019}, needed in the simulation of X-ray absorption \cite{Bokarev_PRL_2013, Pinjari_JCC_2016},
resonant inelastic scattering~\cite{Josefsson_JPCL_2012, Bokarev_JPCC_2015},and photoemission spectra~\cite{Grell_JCP_2015, Golnak_SR_2016, Norell_PCCP_2018a} suggesting its application to \ac{AIS}.
%
%

%
%
Although the ultimate goal is to investigate molecules, this proof-of-concept contribution focuses on the simulation of the prototypical neon $1s^{-1}3p$ \ac{AES} to calibrate the approach, since highly accurate reference data are available from both theory \cite{Stock_PRA_2017} and experiment
~\cite{Kivimaki_JESRP_2001, Yoshida_JPBAMOP_2000, Ueda_PRL_2003, Tamenori_JPBAMOP_2014}.
%
%
Special attention is paid to the representation of the radial continuum waves, which is investigated herein by a thorough test of different approximations.
Note that our implementation allows to calculate molecular \ac{AIS} as well as \ac{PES} which will be presented elsewhere.
%
%

%
We commence this article with an introduction to the underlying theory and further give important details of our implementation.
It continues with the computational details and the benchmark of our results against theoretical and experimental references.
Finally, we conclude the discussion and present perspectives for the molecular application.

%
\section{Theory}
\label{sec_theory}
%
%
%
The approach for the calculation of partial autoionization rates comprises the following approximations:

(i) The two-step model \cite{Wentzel_ZP_1927} is employed, i.e., excitation and decay processes are assumed to be decoupled and interference effects between photoionization and autoionization are neglected, Fig.~\ref{fig:intro}.
%
Within this approximation, the partial autoionization rate for the decay $i \rightarrow \alpha$ reads~\cite{Aberg_CaRiMI_1982}
\begin{equation}
  \label{eq:partialRate}
  \Gamma_{i\alpha}=2\pi\abs{\mel{\Psi_\alpha}{\mathcal{H}-\mathcal{E}_i}{\Psi_i }}^2\,.
\end{equation}
Atomic units are used everywhere, unless explicitly stated otherwise.
%

%
\bt{(ii) We require that all bound state wave functions have the form of \ac{CI} expansions in terms of $N$-electron Slater determinants:}
%
\begin{equation}
 \label{eq:CI}
   \bt{
 \ket{\Psi} = \sum_jC_j \cdot a^\dag_{j_1,\sigma_1}\cdots a^\dag_{j_N,\sigma_N}\ket{0}\,.
      }
\end{equation}
%
\bt{The $a^\dag_{i,\sigma_i}$ are the usual fermionic creation operators in the spin-orbital basis $\{\varphi_{i,\sigma_i}\}$.}
%
%
In this work, multi-configurational bound state wave functions for the unionized and ionized states $\ket{\Psi_i}$ and $\ket{\Psi^+_f}$ are obtained with the \ac{RASSCF} or \ac{RASPT2} method.
However, the presented protocol can employ any \ac{CI}-like \ac{QC} method.
%
%

(iii) 
The limit of weak relativistic effects is assumed, thus the total spins $S,S^+$ and their projections onto the \bt{quantization axis} $M,M^+$ of the bound unionized and ionized system are good quantum numbers.
Further, $S$ and $M$ of the unionized species are conserved during the process.
%
%

%
(iv) The single-channel scattering theory framework is employed, disregarding interchannel coupling, as well as correlation effects between the bound and outgoing electrons.
(v) The continuum orbitals are treated as spherical waves, subject to the spherically averaged potential $V_f(r)$ of the ionic state $\ket{\Psi_f^+}$.
Hence, the continuum orbitals have the form
\begin{equation}
 \label{eq:contOrb}
 \psi_{\alpha,\sigma}(r,\vartheta,\phi) = \frac{1}{r} w^{fk}_{l}(r)Y_{l}^{m}(\vartheta,\phi)\zeta(\sigma) \,,
\end{equation}
with spherical harmonics $Y_l^m(\vartheta,\phi)$ being the angular part and $\zeta(\sigma)$ the spin function.
%
For brevity, $\zeta(\sigma)$ is generally omitted and present only when needed.
The composite channel index $\alpha=(f,l,m,k)$ contains the index of the ionized state $f$, the orbital and magnetic quantum numbers $l$, $m$ and the wave number $k=\sqrt{2\varepsilon_\alpha}$ of the continuum orbital.
This notation uniquely identifies the total energy $\mathcal{E}_\alpha=\mathcal{E}_f + \varepsilon_\alpha$, the continuum orbital and bound state for each channel $\ket{\Psi_\alpha}$.
%
%
Generally, indices $i$ and $f$ always refer to bound states of the unionized and ionized species and $\alpha$ denotes decay channels.
The radial part $w_{l}^{fk}(r)/r$ is determined by solving the radial Schr\"odinger equation
\begin{equation}
 \label{eq:radSE}
 \left(\frac{d^2}{dr^2} + 2\left(\frac{k^2}{2} - V_f(r)\right) - \frac{l(l+1)}{r^2}\right) w^{fk}_{l}(r) = 0\,.
\end{equation}
%

With the assumptions (i)-(v), the $N$-electron ionized continuum states with conserved total spin and projection of the unionized states, $S$ and $M$, can be written as:
\begin{equation}
\label{eq:continuum}
\ket{\Psi_\alpha} = \sum_{M^+=-S^+}^{S^+}\sum_{\sigma=-\frac{1}{2},\frac{1}{2}} C_{S^+,M^+; \sigma}^{S,M}\ket{\Upsilon_{\alpha}^{M^+,\sigma}}\,,
\end{equation}
where $\sigma$ is the spin projection of the outgoing electron.
The Clebsch-Gordan coefficients $C_{S^+,M^+;\sigma}^{S,M} = \braket{S,M}{S^+,M^+;\frac{1}{2}, \sigma}$ couple the channel functions $\ket{\Upsilon_\alpha^{\sigma M^+}}$.
%
%
These are constructed by inserting an additional electron with the continuum orbital $\ket{\psi_{\alpha,\sigma}}$ into the bound ionic state with spin projection $M^+$, retaining the anti-symmetry:
%
\begin{equation}
\label{eq:chanFunc}
\ket{\Upsilon_\alpha^{\sigma M^+}} = a^\dag_{\alpha,\sigma}\ket{\Psi_{f,M^+}^+}\,.
\end{equation}
%
%
%
Note that in contrast to the continuum state $\ket{\Psi_\alpha}$ which is an eigenstate of $\mathbf{S}^2$ and $S_z$, the 
$\ket{\Upsilon_\alpha^{\sigma M^+}}$ are not eigenfunctions of $\mathbf{S}^2$, but only of $S_z$.
%

%
%
%
%
%
%
\section{Details of the implementation}
\label{sec_imp}
%

\subsection{Model potentials}
\label{sec_imp:pot}
%
We commence with the discussion of different approximations to the potential $V_f(r)$ in Eq.~\eqref{eq:radSE} which are central for the quality of the free electron function.
%
%
In this work, we have used the following models:
%
\begin{subequations}
 \begin{alignat}{2}
  \label{eq:pot:free}
  &V^{\text{free}} && = 0\,,\\
  \label{eq:pot:eff}
  &V^{\text{eff}}(r)&& = -\frac{Z_{\text{eff}}}{r}\,,\\
  \label{eq:pot:scr}
  &V_f^{\text{scr}}(r) && = -\frac{Z_f(r)}{r}\,,\\
  \label{eq:pot:dir}
  &V_f^\text{J}(r) && = -\frac{Z}{r} + J_f(r)\,,\\
  \label{eq:pot:dirX}
  &V_f^\text{JX}(r) && =  -\frac{Z}{r} + J_f(r) + X^\text{S}_f(r)\,.
\end{alignat}
\end{subequations}
In words, we use (a) no potential; (b) an effective Coulomb potential with a fixed charge $Z_\text{eff}$; (c) the screened Coulomb potential of an ionic state $f$ with the charge $Z_f(r)$ varying with distance;
\bt{(d) the nuclear, $-Z/r$, and spherically averaged direct potential of the ionic state, $J_f(r)$};
(e) the potential of (d) augmented with Slater's exchange term $X^\text{S}_f(r)$~\cite{Slater_PR_1951}.
\bt{Details on the definition of $Z_f(r)$, $J_f(r)$ and $X^\text{S}_f(r)$ are given in Appendix~\ref{sec:Ap_pot}.}
\subsection{Continuum orbitals}
\label{sec_imp:contOrb}
%
The solutions to the radial Schr\"odinger equation, Eq.~\eqref{eq:radSE}, using the potentials~\eqref{eq:pot:free}--\eqref{eq:pot:dirX} can be understood as follows:
In (a), assuming a free particle, we completely neglect any influence of the ion onto the outgoing electron.
Here, the radial part of $\psi_{\alpha,\sigma}(r,\vartheta,\varphi)$ corresponds to spherical Bessel functions $j_l(kr)$ \cite{Olver2010}:
\begin{equation}
  \label{eq:orbFree}
  \psi^\text{free}_{\alpha,\sigma}(r,\vartheta,\phi) = \bt{\sqrt{\frac{2}{\pi} k}} \cdot  j_l(kr)Y_l^m(\vartheta, \phi) \zeta(\sigma)
\end{equation}
The prefactor \bt{$\sqrt{\frac{2}{\pi}k}$} ensures the correct normalization,
\begin{equation}
  \label{eq:orbNorm}
  \braket{\psi_{\alpha,\sigma}}{\psi_{\alpha',\sigma'}} = \delta_{ll'}\delta_{m_lm_l'}\delta_{\sigma\sigma'} \bt{\delta(\varepsilon_\alpha-\varepsilon_{\alpha'})}\,,
\end{equation}
which is exact only for the free particle approach, and approximate for all other potentials, due to the long range Coulomb distortion.
%

%
An effective Coulomb form (b) is the simplest approach to approximately account for the ionic potential. 
It has the advantage that the solutions to Eq.~\eqref{eq:radSE} are still analytically available in the form of regular Coulomb functions $F_l$ \cite{Olver2010}:
\begin{equation}
  \label{eq:orbEff}
  \psi^\text{eff}_{\alpha,\sigma}(r,\vartheta,\phi) = \bt{\sqrt{\frac{2}{\pi k}}}\cdot \frac{F_l(\eta,kr)}{r}Y_l^m(\vartheta,\phi) \zeta(\sigma)\,,
\end{equation}
with $\eta = -Z_\text{eff}/2k$.
The approaches (c-e) employ numerically obtained potentials according to Eqs.~\eqref{eq:pot:scr}--\eqref{eq:pot:dirX} and thus require a numerical solution of the radial Schr\"odinger equation~\eqref{eq:radSE}.
\bt{
It is done using Numerov's method and the numerically obtained radial waves are scaled such as to satisfy the asymptotic boundary conditions:
}
\begin{subequations}
\begin{align}
  \label{eq:bcZero}
    w_l^{fk}(r\rightarrow 0) &= n r^{l+1}\,,\\
  \label{eq:bcInfty}
    w_l^{fk}(r\rightarrow \infty) &= \bt{\sqrt{\frac{2}{\pi k}}} \Bigl( \cos{\delta^f_l(k)} F_l(\eta, kr ) \nonumber\\
    & \hspace{0.90cm}+ \left. \sin{\delta^f_l(k)} G_l(\eta, kr) \right)\,.
\end{align}
\end{subequations}
The scaling factors and phase shifts $\delta^f_l(k)$ are obtained by matching the numerical solutions and their first derivatives with a linear combination of the regular and irregular Coulomb functions $F_l$ and $G_l$, see Ref.~\citenum{Olver2010}.
This matching is carried out in the asymptotic region where the potential is well approximated by the Coulomb potential of the ions net charge $V_f(r)\approx -Z_\text{net}/r$.
%
%
%

%
Since the screened Coulomb potential model, Eq.~\eqref{eq:pot:scr} is an ad-hoc assumption based on the simple idea that the nuclear charge is screened by the integrated electron density, the applicability of this model needs to be tested.
In turn, (d) and (e), employing the spherically averaged direct Coulomb and exchange terms according to Eqs.~\eqref{eq:pot:dir} and \eqref{eq:pot:dirX}, in a sense correspond to the Hartree and \bt{Slater-$X_\alpha$} levels of accuracy, respectively.
%
%
%

%
%
The continuum orbitals \bt{obtained by any of these methods are not orthogonal to the bound orbitals}, which is in contrast to the behavior that an exact continuum orbital would possess.
\subsection{Continuum matrix elements}
\label{sec_theo:ME}
%
%
%
%
\bt{Using the decomposition of the continuum states in terms of channel functions in Eq.~\eqref{eq:continuum} and separating the Hamiltonian into its one and two electron parts $\mathcal{H}=\sum_u h_u+\sum_{u<v} 1/r_{uv}$, the partial decay rate becomes:}
\begin{align}
 \label{eq:partialRateME}
 \Gamma_{i\alpha} = 2\pi\Biggl| & \sum_{M^+,\sigma} C_{S^+,M^+;\sigma}^{S,M} \left[\sum_u \mel{\Upsilon_\alpha^{\sigma M^+}}{h_u}{\Psi_i } \right.\\
 & \left.+ \sum_{u<v} \mel{\Upsilon_\alpha^{\sigma M^+}}{\frac{1}{r_{uv}}}{\Psi_i }- \mathcal{E}_i\braket{\Upsilon_\alpha^{\sigma M^+}}{\Psi_i }\right]\Biggr|^2\,. \nonumber
\end{align}
\bt{Here, $u$ and $v$ are electron indices, $h_u$ contains the electronic kinetic energy and electron-nuclei attraction terms and $1/r_{uv}$ is the electron repulsion.}
%
The expressions for the overlap, one- and two-electron matrix elements in Eq.~\eqref{eq:partialRateME} are obtained by using L\"owdins Slater determinant calculus~\cite{Lowdin_PR_1955}.
Here one has to take into account that the unionized and ionized bound states are obtained in separate \ac{SCF} calculations.
Consequently, they have different sets of $N_{\rm orb}$ spin-orbitals $\{\varphi_i\}$ and $\{\varphi^+_i\}$ that are not mutually orthogonal.
The spin coordinates are implicitly assumed to be assigned as introduced in Eq.~\eqref{eq:CI}.
Then, the respective creation and annilation operators are $a_i^\dag,\,a_i$ and $({a^+_i})^\dag,\,a^+_i$.
With this, the overlap integral in Eq.~\eqref{eq:partialRateME} can be rearranged into the overlap of the continuum orbital and the \ac{DO} $\ket{\Phi_{i\alpha}^{M^+}}$,
\begin{equation}
 \label{eq:ME_overlap}
 \braket{\Upsilon_\alpha^{\sigma M^+}}{\Psi_i} = \braket{\psi_{\alpha,\sigma}}{\Phi_{i\alpha}^{M^+}}\,.
\end{equation}
The \ac{DO} is generally defined as the $N-1$ particle integral over the transition density of the unionized and ionized states $\ket{\Psi_i}$ and $\ket{\Psi_{f,M^+}^+}$ that are associated with the channel $\alpha$
%
and can be expressed in second quantization form as a linear combination of the spin-orbitals $\{\varphi_s\}$ of the unionized species, with the coefficients $\phi^{M^+}_{\alpha,s}$
\begin{equation}
 \label{eq:DO}
 \ket{\Phi_{i\alpha}^{M^+}} = \sum_{s=1}^{N_\text{orb}}\underbrace{{\mel{\Psi_{f,M^+}^+}{a_s}{\Psi_i}}}_{\displaystyle \phi^{M^+}_{\alpha,s}}\ket{\varphi_s}\,.
\end{equation}
\bt{Staying on this route, the one- and two-electron transition matrix elements in Eq.~\eqref{eq:partialRateME} can be expressed as:
\begin{align}
  \label{eq:ME-one-NO}
  \sum_u \mel{\Upsilon_{\alpha}^{M^+,\sigma}}{h_u}{\Psi_i } & = \mel{\psi_{\alpha,\sigma}}{h}{\Phi_{i\alpha}^{M^+}} + \braket{\psi_{\alpha,\sigma}}{\tilde{\Phi}_{i\alpha}^{1, M^+}}
  %
\end{align}
and
\begin{align}
  \label{eq:ME-two-NO}
  \sum_{u<v}\mel{\Upsilon_\alpha^{\sigma M^+}}{\frac{1}{r_{uv}}}{\Psi_i} & = \sum_{q=1}^{N_{\rm orb}} \mel{\psi_{\alpha,\sigma}\varphi^+_{q}}{\frac{1}{r_{12}}}{\Xi_{i\alpha}^{M^+,q}} \\
  & + \braket{\psi_{\alpha,\sigma}}{\tilde{\Phi}_{i\alpha}^{2, M^+}}\,. \nonumber
\end{align}
Here, $\ket{\Xi_{i\alpha}^{M^+,q}}$ is the two-electron reduced transition density
\begin{equation}
  \label{eq:2DO}
  \ket{\Xi_{i\alpha}^{M^+,q}} = \sum_{s_1\neq s_2}^{N_{\rm orb}}\mel{\Psi_{f,M^+}^+}{(\hat{a}_{q}^+)^\dag\hat{a}_{s_1}\hat{a}_{s_2}}{\Psi_i} \ket{\varphi_{s_1}\varphi_{s_2}}\,,\\
\end{equation}
and $\ket{\tilde{\Phi}_{i\alpha}^{n,M^+}}$ are the one- and two-electron conjugated Dyson orbitals for $n=1$ and 2, respectively, see Eqs.~\eqref{eq:con1DO} and~\eqref{eq:con2DO}.
%
%
A simpler formulation of the NO terms proposed in Ref.~\citenum{Manne_CP_1985} is not used herein, because in practice it is not always strictly equivalent to our approach, check \supp~Section II for details.
}
%

%
%
\bt{The \ac{SO} approximation, i.e., the assumption that the overlaps of the continuum and all unionized orbitals are zero, $\braket{\psi_{\alpha,\sigma}}{\varphi_{i,\sigma_i}} = 0$, implies that}
\bt{the overlap integrals between the continuum and the ordinary and conjugated Dyson orbitals in Eqs.~\eqref{eq:ME_overlap},~\eqref{eq:ME-one-NO}, and~\eqref{eq:ME-two-NO} vanish.}
\bt{Since the evaluation of the conjugate \acp{DO} can be quite involved,
 \ac{SO} approximation substantially simplifies the computation.}

%
\bt{A similar effect is achieved by using the \ac{GS} orthogonalization to project the orbitals of the ionized system out of the continuum functions and enforce $\braket{\psi^{\text{GS}}_{\alpha,\sigma}}{\varphi^+_{i,\sigma_i}} = 0$.}
%
This approach has been tried before, e.g., in Ref.~\citenum{Zahringer_PRA_1992} and will be tested herein as well.
In the following, we will use the labels \ac{SO}, NO and \ac{GS} to indicate that the
\bt{overlap terms in Eqs.~\eqref{eq:ME_overlap},~\eqref{eq:ME-one-NO}, and~\eqref{eq:ME-two-NO} have been neglected (\ac{SO}), fully included (NO), and that \ac{GS}-orthogonalized continuum functions have been used to evaluate the partial rates including the NO terms as well.}
Further, we also compare the results of the ``full'' Hamiltonian coupling, against the popular choice to account only for the two-electron terms in Eq.~\eqref{eq:partialRateME}, denoting them as $\mathcal{H}$  and $r^{-1}$ coupling, respectively.
\bt{The combination of matrix elements corresponding to each approach is detailed in Appendix~\ref{sec:Ap_TME}. }

%
To sum up, the present article reports on the influence of different combinations of the introduced approximations to the potentials, transition matrix elements, and continuum orbitals on the partial Auger decay rates of the exemplary neon $1s^{-1}3p$ resonance.
As a shorthand notation for the combination of the different approximations we will use the $coupling$~$\sep$~$potential$~$\sep$~$nonorthogonality$ notation, were applicable.
For instance, $\mathcal{H} \sep V_f^\text{JX}(r) \sep \text{NO}$ denotes partial rates obtained using the $\mathcal{H}$ coupling, with radial waves corresponding to the $V_f^\text{JX}(r)$ potential including the
\bt{overlap terms}.
%

%
\section{Computational Details}
\label{sec_comp}
%
\subsection{Quantum chemistry for bound states}
\label{sec_comp:QC}
The wave functions of the bound neutral and ionic states have been obtained in separate state-averaged \ac{RASSCF} calculations with a locally modified version of MOLCAS 8.0 \cite{Aquilante_JCC_2016}.
To prevent mixing of different angular momentum basis functions into one orbital, the "atom" keyword has been employed.
The \ac{QC} schemes used to evaluate the bound states $\ket{\Psi_i}$ and $\ket{\Psi_f^+}$ are presented in Table~\ref{tab:QC}.
%
%
\begin{table}[htp]
  \begin{tabular}{l| r | r | r | r }
         &             &          & \multicolumn{1}{c|}{States} & \\
    QC   & Basis Set   & \ac{AS}  & Ne/Ne${}^+$                  & \acs{X2C}\\
    \hline
    I        & [7s6p3d2f]       & \ac{RAS}$(8;1,1)$     & 41/132    & no \\
    II       & [22s6p3d2f]-rcc  & \ac{RAS}$(8;1,1)$     & 41/132  & yes \\
    III       & [9s8p5d4f]       & \ac{RAS}$(26;1,1)$    & 131/420   & no \\
  \end{tabular}
  \caption{\label{tab:QC} \ac{QC} setups and number of states used in the state-averaged \ac{RASSCF} calculations}
\end{table}
The \ac{RAS} formalism is a flexible means to select electronic configurations.
Therein, the \ac{AS} is subdivided into three subspaces RAS1, RAS2 and RAS3.
The \ac{RAS} notation that is used throughout the paper is to be understood as follows:
In all spaces, the $1s$ orbital forms the \ac{RAS}1 subspace and the $2s$ and $2p$ orbitals build up the RAS2 one.
The occupation of the $2s$ and $2p$ orbitals in RAS2 is not restricted, while only $h$ electrons may be removed from RAS1.
Finally the RAS3 subspace contains $v$ virtual orbitals that can be occupied by at most $p$ electrons.
Thus, we can herein uniquely specify each \ac{AS} as RAS$(v;h,p)$.
The \acp{AS} used in this study are: \ac{RAS}$(8;1,1)$, containing $3s$, $3p$, $4s$, and $4p$ orbitals in the RAS3;  \ac{RAS}$(26;1,1)$, enlarging RAS3 by the $3d$, $4d$, $5s$, $5p$, $6s$, and $6p$ orbitals and
\ac{RAS}$(33;1,1)$, adding the $4f$ orbitals.
The number of configurations possible with each \ac{AS} is shown in Table \ref{tab:CSFs}.
\begin{table}[htp]
  \begin{tabular}{l|c|c|c}
    \ac{RAS} & \ac{RAS}$(8;1,1)$ & \ac{RAS}$(8;26,1)$ & \ac{RAS}$(8;33,1)$\\
    \hline
    Ne        & 41  & 131 & 166 \\
    Ne${}^+$  & 197 & 629 &  797 \\
  \end{tabular}
  \caption{\label{tab:CSFs} Maximum number of configurations for each \ac{AS}}
\end{table}
For all \ac{QC} schemes (see Table~\ref{tab:QC}) all states are included in the \ac{RASSCF} procedure for the Ne wave functions, while the core excited states have been excluded for the calculations of Ne${}^+$.
%

%
%
\ac{ANO} type basis sets have been employed using a $(22s17p12d11f)$ primitive set.
It was constructed by supplementing the \ac{ANO} exponents for neon \cite{Widmark_TCA_1990} in each angular momentum with eight Rydberg exponents generated according to the scheme proposed by Kaufmann et al. \cite{Kaufmann_JPBAMOP_1989}.
The contractions were then obtained with the GENANO module \cite{Almlof_JCP_1987} of OpenMolcas~\cite{Galvan_C_2019} using density matrices from state averaged \ac{RASSCF} calculations for Ne and Ne${}^+$ with the \ac{RAS}$(33;1,1)$ \ac{AS}.
All possible 166 states for Ne have been taken into account but for Ne${}^+$, the core excited manifold has been excluded, leading to 532 states.
To get the final basis set, the sets obtained for  Ne and Ne${}^+$ have been evenly averaged.
In this work, we use the contractions: [7s6p3d2f], obtained by the procedure described above; [9s8p5d4f], corresponding to [7s6p3d2f] supplemented with Rydberg contractions that resulted from the GENANO procedure using the "rydberg" keyword; [22s6p3d2f]-rcc, similiar to [7s6p3d2f] but with uncontracted $s$ functions and scalar relativistic corrections according to the \ac{X2C} scheme \cite{Peng_TCA_2012}\bt{, see \supp~Sec.~III}.
%

%
%
All energies have been corrected using the single state \ac{RASPT2} method \cite{Malmqvist_JCP_2008b} with an imaginary shift of $0.01$~a.u.
%
%
\bt{To be consistent with the basis set generation, scalar relativistic effects have only been included for \ac{QC} II.}
%
%
The \ac{RASSI} \cite{Malmqvist_CPL_2002a} module of MOLCAS was used to compute the biorthonormally transformed orbital and \ac{CI} coefficients \cite{Malmqvist_IJQC_1986} for the atomic and ionic states: $\{\varphi_i\}, \{C_j\} \rightarrow \{\tilde{\varphi}_i\},
 \{\tilde{C}_j\}$ such that $\braket{\tilde{\varphi}^+_i}{\tilde{\varphi}_j} = \delta_{ij}$, while the total wave functions remain unchanged.
This biorthonormal basis is used in the evaluation of the Dyson orbitals, Eqs.~\eqref{eq:DO},
\bt{\eqref{eq:con1DO},~\eqref{eq:con2DO}},
and two-electron reduced transition densities, Eq. \eqref{eq:2DO}.
%
\subsection{Matrix elements in the atomic basis}
\label{ses_comp:ME}
%
%
The matrix elements between bound orbitals occuring in Eqs.\eqref{eq:partialRateME}, \eqref{eq:ME-one-NO}, and \eqref{eq:ME-two-NO} are evaluated by transforming the orbitals to the atomic basis and calculating the atomic basis integrals with the libcint library \cite{Sun_JCC_2015}.
The most time consuming part in the computation of the partial decay rates using Eq.~\eqref{eq:partialRateME} is the estimation of the two-electron continuum-bound integrals.
%
%
Transforming the two-electron reduced transition densities and the orbitals to the atomic basis $\{\chi_i\}$ and neglecting the spin integration, the two-electron continuum-bound integrals 
used in the practical evaluation of Eq.~\eqref{eq:ME-two-NO} read as:
\begin{equation}
  \label{eq:2ME-atomic}
  \braket{\alpha a}{b c} = \int \psi^*_\alpha(\mathbf{r}_1) \chi_b(\mathbf{r}_1) \underbrace{\left [ \int \frac{\chi_a^*(\mathbf{r}_2) \chi_c(\mathbf{r}_2)}{r_{12}})  d\mathbf{r}_2^3 \right]}_{f_{ac}(\mathbf{r}_1)} d\mathbf{r}^3_1 \,.
\end{equation}
The function $f_{ac}(\mathbf{r}_1)$ is similar to an atomic nuclear attraction integral, and is evaluated as a function of $\mathbf{r}_1$ using the libcint library \cite{Sun_JCC_2015}.
An important point is to exploit the fact that the kinetic energy of the continuum electron is only encoded in its radial part.
Transforming to spherical coordinates $\mathbf{r} \rightarrow (r,\Omega)$ centered at the origin of the outgoing electron allows to separate off the radial integration:
\begin{align}
  \label{eq:2ME-atomic-radial}
  \braket{\alpha a}{b c} & =  \int_0^\infty r w_l^{fk}(r) \times \nonumber\\
                &\times \underbrace{\left[ \int_{\Omega}Y_l^m(\Omega)\chi_b(\mathbf{r}(r,\Omega)) f_{ac}(\mathbf{r}(r,\Omega)) d\Omega\right ]}_{F_{abc}(r)} dr\,.
\end{align}
%
%
The angular integral in $F_{abc}(r)$ is determined numerically by using the adaptive two-dimensional integration routine \texttt{cuhre} from the Cuba 4.2 library \cite{Hahn_CPC_2005a}.
To reduce the number of points at which $F_{abc}(r)$ is evaluated, an adaptive spline interpolation is used, which was developed by us and implemented in our code.
Therein, the grid spacing is adjusted such that the absolute error estimate of the interpolation
is kept lower than $10^{-6}$~a.u. on each region with a different spacing. 
Note that the $F_{abc}(r)$ are determined only once, while the final radial integration in Eq.~\eqref{eq:2ME-atomic-radial} needs to be evaluated for every transition $i\rightarrow \alpha$.
The radial integration in Eq.~\eqref{eq:2ME-atomic-radial} is carried out using the Simpson rule.
For the one-electron continuum-bound integrals that are needed in the evaluation of the one-electron matrix elements in Eq.~\eqref{eq:partialRateME}, an analogous approach has been implemented.
\bt{This protocol has been developed aiming for the application to molecular systems.
Hence, it does not exploit that the atomic orbitals are eigenstates of angular momentum operator, which would be usual in a purely atomic approach.}

%
\section{Results and Discussion}
\label{sec_res_disc}
%
Here, we perform a thorough benchmark of the approaches to evaluate \ac{AES} presented in Section~\ref{sec_theory} on the exemplary Auger decay of the neon $1s^{-1}3p$ resonance.
First, in Section~\ref{sec_res_disc:theo}, we compare the \ac{AES} modelled with our protocol against data obtained from an atomic \ac{MCDF} calculation~\cite{Stock_PRA_2017}, which serves as a high level theoretical reference.
Second, in Section~\ref{sec_res_disc:exp}, we undertake the comparison to experimental results.
The comparison against both theory and experiment is needed since no uniform and highly resolved experimental data covering the full energy range discussed herein has been published to date.
%
%
\subsection{Benchmark of theoretical models}
\label{sec_res_disc:theo}
%
\begin{figure*}
  \begin{minipage}{\textwidth}
     \includegraphics{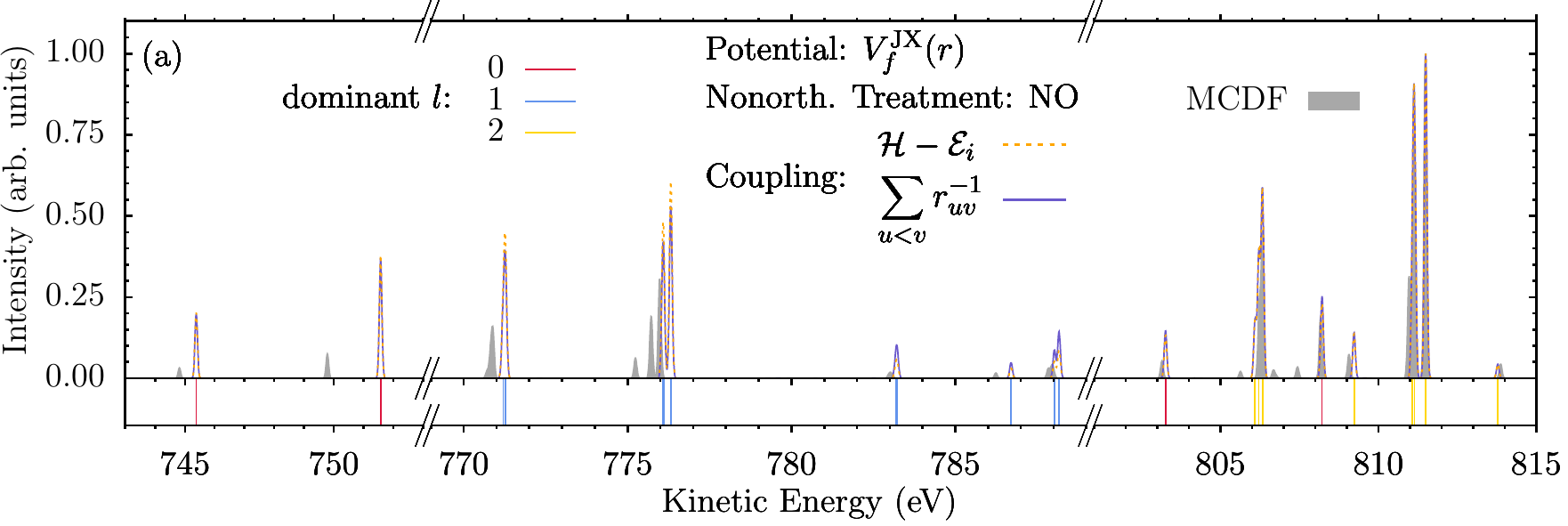}
  \end{minipage}
  \begin{minipage}{\textwidth}
     \includegraphics{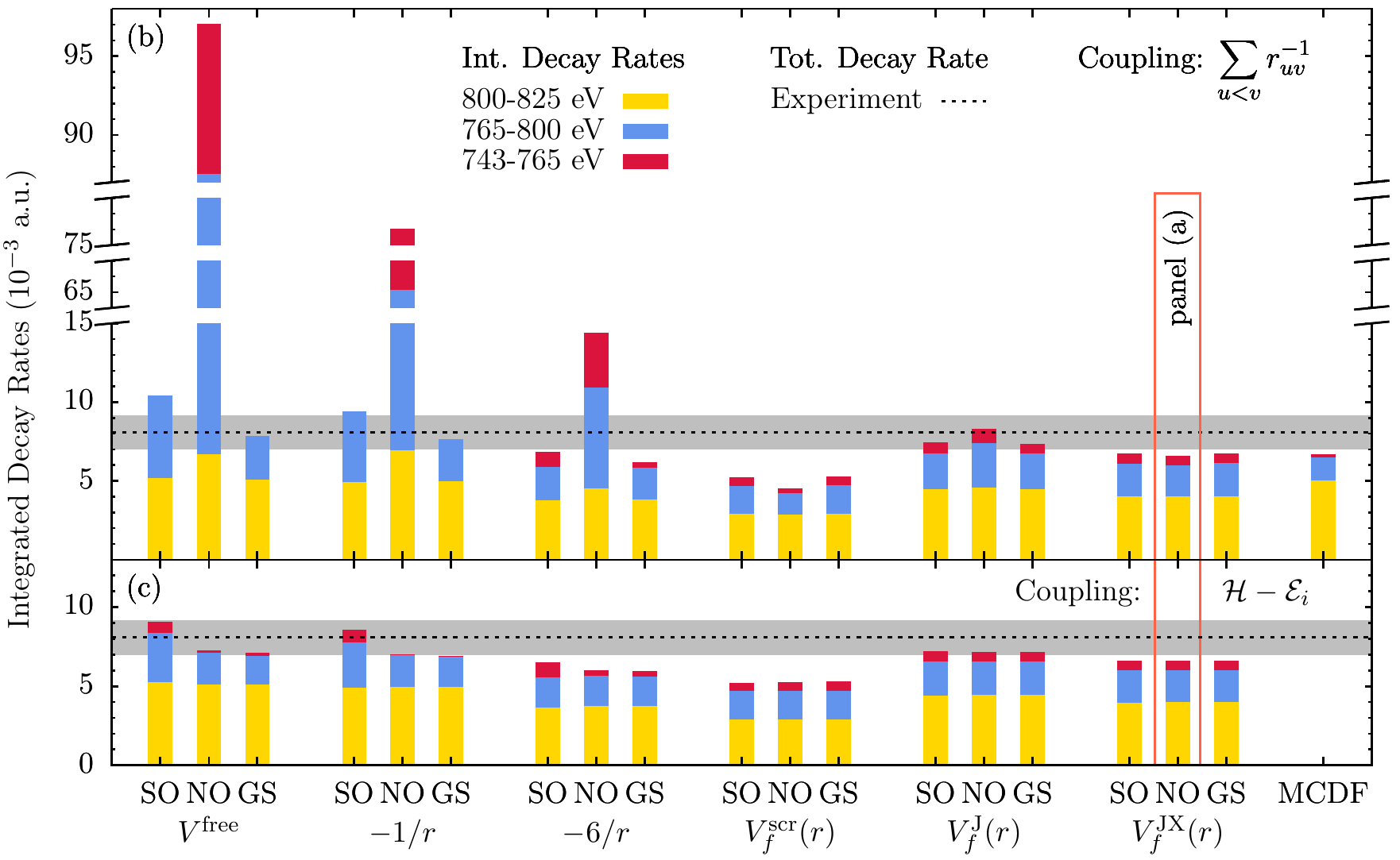}
  \end{minipage}
  \caption{\label{fig:res:QC1-normInt} (a) Neon $1s^{-1}3p$ \ac{AES} obtained with the $r^{-1}$ and $\mathcal{H}$ couplings and the NO approach using continuum orbitals generated by the $V_f^\text{JX}(r)$ potential are shown in comparison to the \ac{MCDF} results reported by Stock et al.~\cite{Stock_PRA_2017}.
  All spectra are broadened using a Gaussian profile with an FWHM of $0.1$~eV, normalized to the peak at $811.5$~eV, and shifted globally to align the $811.5$ eV peak with experimental data \cite{Kivimaki_JESRP_2001}.
  The spectra from (a) correspond to the NO histograms of the $V_f^\text{JX}(r)$ potential in panels (b) and (c).
  %
  The vertical lines at the bottom of panel (a) indicate the predominant continuum orbital angular momentum ($l$) contribution to each peak.
  (b) \& (c)  Auger decay rates integrated over the given energy ranges corresponding to distinct $l$ contributions.
  The decay rates have been evaluated based on \ac{QC} model I for the $r^{-1}$ (b) and $\mathcal{H}$ couplings (c).
  The radial continuum functions correspond to the depicted potentials.
  Nonorthogonality of the continuum and bound orbitals was treated with the \ac{SO}, NO, and \ac{GS} approaches, see text.
  %
  The \ac{MCDF} data have been scaled such that the total decay rate matches the one obtained for the $r^{-1} \sep V_f^\text{JX}(r) \sep \text{NO}$ approach.  %
  %
  For reference, the experimentally   determined total rate of $8.08\pm1.1\times 10^{-3}$~a.u. ($0.22\pm0.03$~eV) \cite{Avaldi_PRA_1995} 
  is depicted as a horizontal dashed line.
  %
  The gray region indicates the experimental uncertainty.
          }
\end{figure*}
%
%
Panel (a) of Fig.~\ref{fig:res:QC1-normInt} shows neon \acp{AES}, resulting from the autoionization of the $1s^{-1}3p$ states, obtained using bound state wave functions from \ac{QC} scheme I (cf. Table~\ref{tab:QC}) with radial waves corresponding to the spherically averaged direct-exchange potential $V^\text{JX}_f(r)$ Eq.~\eqref{eq:pot:dirX}.
The partial rates have been evaluated using the full $\mathcal{H}$ coupling as well as the approximate $r^{-1}$ coupling in Eq.~\eqref{eq:partialRateME}.
Further, the nonorthogonality of the continuum and bound orbitals was accounted for by including the
\bt{overlap terms, NO, see Eqs.~\eqref{eq:ME_overlap},~\eqref{eq:ME-one-NO}, and \eqref{eq:ME-two-NO}}.
A spectrum employing $r^{-1}$ coupling at the \ac{MCDF} level, obtained by Stock et al. \cite{Stock_PRA_2017} with the RATIP package \cite{Fritzsche_CPC_2012},  serves as a theoretical benchmark.
Therein, the atomic structure was obtained with a configuration space including single electron excitations from the $1s$, $2s$, and $2p$ to the $np$ up to $n=7$ and $3d$ orbitals.
The four-component continuum orbitals have been obtained as distorted waves within the potential of the respective ionized atomic state.
%

%
%
All spectra have been constructed by assigning a Gaussian lineshape with an FWHM of $\gamma=0.1$~eV to each channel.
\begin{equation}
  \label{eq:AES}
  \mathrm{AES}_{1s^{-1}3p}(\varepsilon) = \sum_\alpha \Gamma_{1s^{-1}3p \,\alpha} G(\varepsilon - \varepsilon_\alpha,\gamma)
\end{equation}
Where $ G(\varepsilon,\gamma)=\sqrt{\ln{2}/(\pi\gamma)}\,\, \mathrm{exp}(-{4\ln{2}} \varepsilon^2 /{\gamma^2})$, $\varepsilon$ is the kinetic energy of the emitted electrons and $\varepsilon_\alpha = \mathcal{E}_{1s^{-1}3p} - \mathcal{E}_f$ is the Auger electron energy of
the channel $\ket{\Psi_\alpha}$.
The spectra were normalized to the peak height at $811.5$~eV, and shifted globally by $-5.35$~eV (\ac{QC} I) and $-2.45$~eV (\ac{MCDF}) such that the peak at $811.5$~eV is aligned to the experimental data taken from Kivim\"aki et al. \cite{Kivimaki_JESRP_2001} (see Fig.~\ref{fig:res:exp}).
In addition, the dominant continuum orbital angular momentum contribution to the intensity is indicated for each peak.
This shows that the regions $743-765$~eV,  $765-800$~eV and $800-825$~eV correspond almost exclusively to the emission of $s$, $p$ and $d$ waves, respectively,
\bt{with the exception that the peaks at $803$~eV and $808$~eV lie in the $d$ region but are due to $s$ wave emission.}
Hence, we will refer to this regions as $s$, $p$, $d$ rather than using the energies in what follows.
%

%
\bt{
It is evident from Fig.~\ref{fig:res:QC1-normInt} (a) that the normalized \ac{AES} obtained for the $\mathcal{H}$ and $r^{-1}$ couplings are almost indistinguishable in the $s$ and $d$ regions.
In contrast, the features in the $p$ region at $771$~eV and $776$~eV are enhanced by $10\%$, while the small bands in the range $783-788$~eV, are reduced by about $40\%$, if the $\mathcal{H}$ coupling terms are included.
The overall effect on the spectrum, however, is still rather small and neglecting the contributions beyond the $r^{-1}$ coupling in the partial rate evaluation seems to be justified in this case.
}
The comparison with the \ac{MCDF} data in the $d$ region shows that the Auger energies and relative intensities are overall quite well reproduced by our approach.
Only the intensities of two small satellite peaks at $803$~eV and $809$~eV are overestimated and three tiny features around $805.5-807.5$~eV are not present when using our approach.
The latter deficiency can be safely attributed to the smaller configuration space employed in the \ac{QC} I scheme as compared to the one used to obtain the \ac{MCDF} results in Ref.~\citenum{Stock_PRA_2017}.
Looking at the $s$ and $p$ regions, the Auger energies from the \ac{QC} model I become slightly but increasingly blue shifted with respect to the \ac{MCDF} ones at the lower energy flank of the spectrum.
The intensities in turn are considerably overestimated by factors of about five and two for the $s$ and $p$ regions, respectively. 
Note that very similiar spectra are obtained if the \ac{RASPT2} energy correction is not used (\supp~Fig.~S1)
Panels (b) and (c) of Fig.~\ref{fig:res:QC1-normInt} show the integrated decay rates for the $s$, $p$ and $d$ regions evaluated using QC model I with the $r^{-1}$ (b) and $\mathcal{H}$ couplings (c), comparing different approaches to compute the partial decay rates.
\bt{The respective spectra are given in \supp~Figs.~S2-S7.}
The rates were obtained for all combinations of potential models in Eqs.~\eqref{eq:pot:free} -~\eqref{eq:pot:dirX} with the different nonorthogonality approaches \ac{SO}, NO, and \ac{GS} for the continuum orbitals.
Since no absolute rates for the \ac{MCDF} results are available, these have been scaled to the total decay rate obtained using  the $r^{-1} \sep V_f^\text{JX}(r) \sep \text{NO}$ treatment.
The data in panels (b) and (c) show that the decay rates corresponding to the $s$, $p$ and $d$ spectral regions converge for both couplings as the quality of the potential is increased from the free particle approximation $V^\text{free}$ to the spherically averaged direct-exchange potential $V_f^\text{JX}(r)$.
The $s:p:d$ - ratio obtained from the \ac{MCDF} spectrum, however, is not matched.
Our approaches systematically overestimate the decay rates due to the $s$ and $p$ channels, in line with the mismatch of intensities unveiled in panel (a).
\bt{
Prominently, using the approximate $r^{-1}$ coupling together with the NO overlap terms leads to an overestimation of the $s$ and $p$ regions with respect to the \ac{SO} results by an order of magnitude for $V^\text{free}$ and $-1/r$, and by factors of $4$ and $3$ for $-6/r$.
In contrast, employing the \ac{SO} approximation or \ac{GS} orthogonalized continuum orbitals results in more realistic decay rates.
Here, the \ac{GS} rates reproduce the \ac{SO} ones for the $d$ region but are considerably smaller in the $s$ and $p$ regions.
However, when either of the more physically sound potentials $V_f^\text{scr}(r)$,  $V_f^\text{J}(r)$, or  $V_f^\text{JX}(r)$ is used, the choice of the nonorthogonality treatment has only a weak influence on the integrated decay rates.
This characteristic is due to the continuum-bound orbital overlaps that decrease if more realistic potentials are chosen, causing the NO overlap terms in Eqs.~\eqref{eq:ME_overlap},~\eqref{eq:ME-one-NO}, and~\eqref{eq:ME-two-NO}, as well as the effect of the \ac{GS}-orthogonalization to become negligible.
Adding to this, the general insensitivity of the $d$ region to the nonorthogonality treatment is explained by the fact that the overlap of the $d$ continuum waves with the bound orbitals vanishes independently of the potential model.
The overlap integrals are in \supp~Table~S1.
}
\bt{
Generally, the inclusion of the $\mathcal{H}$ coupling, Fig.~\ref{fig:res:QC1-normInt} (c), yields mostly smaller total decay rates and less pronounced differences between the \ac{SO}, NO and \ac{GS} results.
The \ac{SO} decay rates are usually the largest when $\mathcal H$ coupling is applied, while the NO and \ac{GS} ones are smaller, underestimating the $s$ and $p$ regions for $V^\text{free}$, $-1/r$, and $-6/r$, with respect to \ac{SO}.
Noteworthy, the huge overestimation due to the combination of $r^{-1}$ coupling and NO terms is mitigated completely.
The reason is that the NO contribution to the matrix element is determined by cancellation effects between different one- and two-electron NO terms rather than their individual magnitude.
Neglecting the one-electron contributions in $r^{-1}$ coupling prohibits this error cancellation, leading to the immense overestimation observed for $V^\text{free}$, $-1/r$, and $-6/r$.
Similar, the agreement between the \ac{GS} and NO results for these potentials is not because the effect of the \ac{GS}-orthogonalization is minute, but again due to the cancellation.
}
%

%

\begin{figure*}
  %
  \begin{minipage}{\textwidth}
   \includegraphics{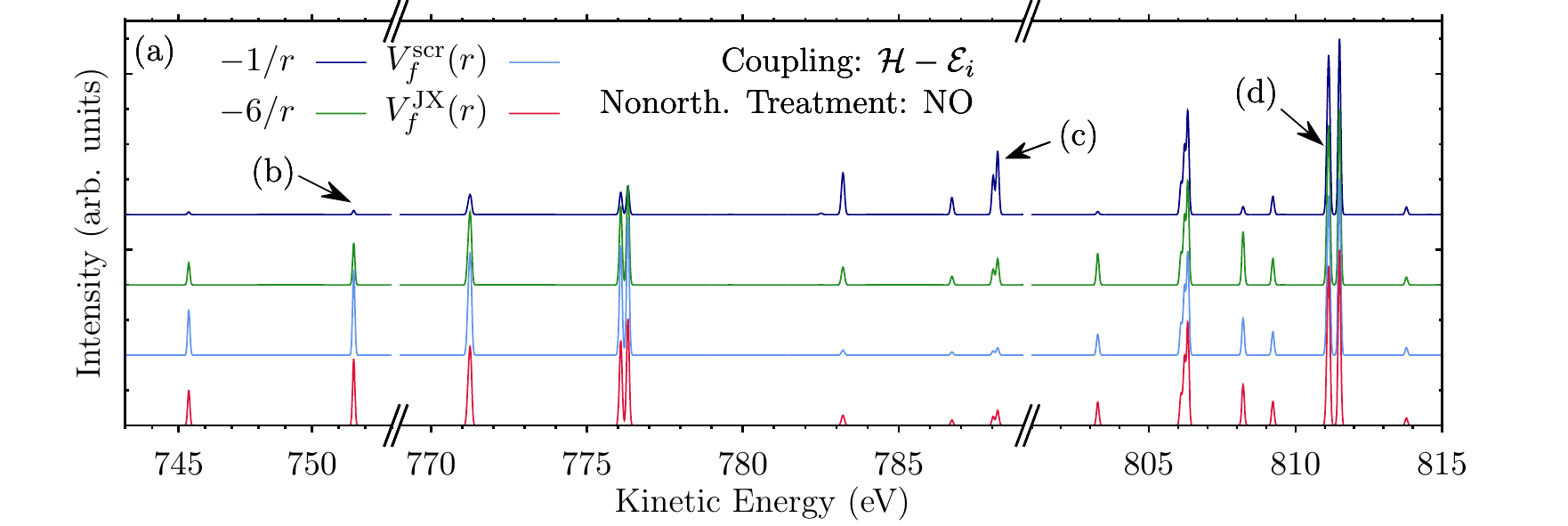}
  \end{minipage}
  \begin{minipage}{\textwidth}
    \includegraphics{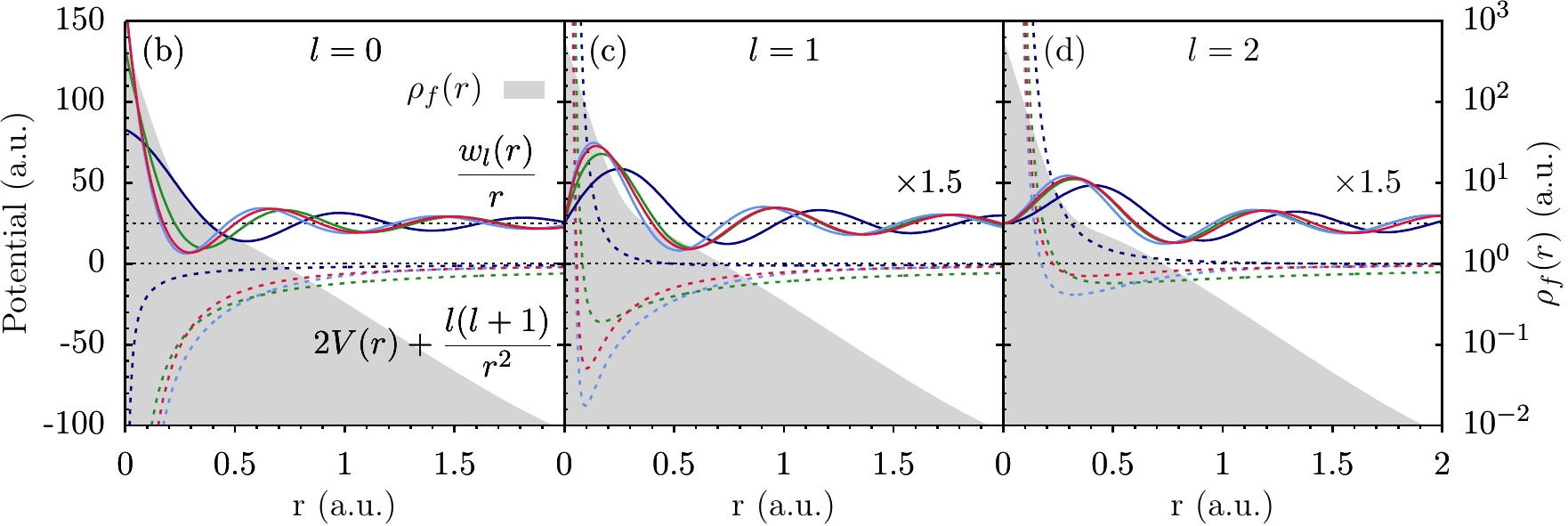}
  \end{minipage}
  \caption{\label{fig:res:QC1-waves} (a) Neon $1s^{-1}3p$ \ac{AES} evaluated using QC scheme I, $\mathcal{H}$ coupling, and the NO nonorthogonality treatment with radial continuum functions corresponding to the given potentials.
  The spectra are normalized to the peak at $811.5$~eV, broadened using a Gaussian FWHM of $0.1$~eV and shifted globally by $-5.35$~eV.
  \bt{In addition, they have have been shifted vertically by $2.0$ ($V_f^\text{scr}(r)$), $4.0$ ($-6/r$), and $6.0$ ($-1/r$) units to enhance the visibility.}
  %
  %
  %
  (b-d) Effective radial potentials (dashed) and continuum functions (solid) corresponding to the dominant angular momentum contribution $l=0,1,2$ of the peaks (b-d) in panel (a) are shown together with the spherically averaged electron densities $\rho_f(r)$ of the respective ionized states.
  The colors of the radial waves and potentials correspond to the spectra shown in panel (a).
  %
          }
\end{figure*}

%
We conclude the discussion of this figure with the observation that the total rates obtained with the $V_f^\text{J}(r)$, and $V_f^\text{JX}(r)$ potentials reproduce the experimentally observed decay rate \cite{Avaldi_PRA_1995}.
%
\bt{Independent of the chosen approach, the best match is obtained with the $V_f^\text{J}(r)$ potential, whereas the $V_f^\text{JX}(r)$ leads to slightly underestimated total rates, which can be attributed to the inclusion of the attractive exchange term, Eq.~\eqref{eq:potX}}.
%
%
Similarly, we assign the considerable underestimation of the total decay rates by the $V_f^\text{scr}(r)$ potential to the fact that it is more attractive in the core region than $V_f^\text{JX}(r)$, see Fig.~\ref{fig:res:QC1-waves}.
\bt{
The $-6/r$ potential yields total rates and spectra that are in approximate agreement with those obtained for $V_f^\text{J}(r)$ and $V_f^\text{JX}(r)$ apart from the case when $r^{-1}$ coupling and the NO terms are used (spectra in \supp~Figs.~S2-S7).
Notably, the total rates obtained in $\mathcal{H}$ coupling with the effective potentials $V^\text{free}$ and $-1/r$ agree well with the experimental reference.
However, despite the good agreement in the total rates the spectra for these approaches deviate in the $p$-region considerably from the ones obtained for the more accurate potentials (\supp~Figs.~S5-S7), indicating that this agreement is rather accidental.
}
%

%
%
%

%
%
%
\bt{
To shed light on the influence of different potentials on the total decay rates and spectra obtained with \ac{QC} model I, $\mathcal{H}$ coupling, and the NO approach, potentials, radial continuum functions and spectra are presented in Fig.~\ref{fig:res:QC1-waves}.
}
Panel (a) contains the normalized \acp{AES} obtained using the potentials $-1/r$, $-6/r$, $V_f^\text{scr}(r)$, and $V_f^\text{JX}(r)$.
%
%
\bt{The spectra have been shifted vertically for the sake of clarity.}
Shifts and broadening parameters are the same as in Fig.~\ref{fig:res:QC1-normInt} (a).
The $d$ region is represented very well with all potential models, with the exclusion that the satellite bands at $803$~eV and $808$~eV are barely present when using the  $-1/r$ potential.
In contrast, the relative intensities of the $s$ and $p$ regions of the spectra are strongly affected by the choice of the potential.
Here, the $p$ region comprises two peak groups with different character.
\bt{
%
Using the $V_f^\text{JX}(r)$ results as a reference for this discussion, the peaks around $771$~eV and $776$~eV are overestimated by $30\%$ with $V_f^\text{scr}(r)$, almost reproduced with $-6/r$  and underestimated by a factor of four when using $-1/r$.
In contrast, the bands between $783$~eV - $788$~eV behave in an inverse manner.
Namely, they are underestimated by $50\%$ with $V_f^\text{scr}(r)$ and overestimated by $50\%$ and a factor of four with $-6/r$ and $-1/r$, respectively.
This behavior of the screened and the effective Coulomb potentials, is caused by the fact that $V_f^\text{scr}(r)$ is more attractive and $-1/r$ as well as $-6/r$ -- less attractive than $V_f^\text{JX}(r)$, as depicted in panels (b)-(d).
This trend is solely dictated by the potential but not the coupling or nonorthogonality treatment; note the exception of $r^{-1} \sep \text{NO}$ combination (\supp~Figs.~S2-S7).
Finally in the $s$ region, the tighter potential results in larger intensities: $-1/r$ and $-6/r$ lead to an underestimation by an order of magnitude and $35\%$, respectively, whereas $V_f^\text{scr}(r)$ yields an overestimation by $25\%$.
%
}
For each region, the radial waves and respective potentials, including the angular momentum term, are shown in the panels (b)-(d) corresponding to the characteristic peaks denoted in panel (a).
%
%
\bt{
Note that the $V^\text{free}$ and $V_f^\text{J}(r)$ cases are very similar to the $-1/r$ and $V_f^\text{JX}(r)$ ones, respectively, and have been excluded from the discussion of this figure.
The respective spectra are in \supp~Fig.~S7.
}
One might understand the differences in the sensitivity of the $s$, $p$ and $d$ spectral regions to the potential by comparing the short range behavior of the radial waves $w_l^{fk}(r)/r$, Eq.~\eqref{eq:bcZero},
with the electron density that is sharply peaked in the core region.
This suggests that the matrix elements in Eq.~\eqref{eq:partialRateME} are very sensitive to the description of the continuum orbitals in this region.
It is well known that only the $s$ waves have a considerable contribution at the core, since the radial functions tend to zero as $r^l$.
In fact, the effective radial potential at the core is dominated by the angular momentum term for $l>0$, meaning that the influence of the present potential models on the total rates should decrease with increasing $l$.
%
%
%
\bt{Regarding the relative changes of the integrated decay rates within each region, this is true for $\mathcal{H}$ and $r^{-1}$ coupling, when the NO overlap terms are included, irrespective of whether the \ac{GS} orthogonalization is used in addition or not, Fig~\ref{fig:res:QC1-normInt} (b) and (c).
However, under the \ac{SO} approximation for $r^{-1}$ couplings, the $p$ spectral region is more sensitive to variations in the potential model than the $s$ region.
}
In fact, for the $s$ waves it is seen that only the potentials $V^\text{scr}_f(r)$ and $V_f^\text{JX}(r)$ lead to similar radial waves, the slight differences being due to the fact that the screening model is too attractive in the core region.
Notably, these slight deviations lead to an underestimation of the total decay rates obtained with the $V^\text{scr}_f(r)$ potential by about $25\%$ in comparison to those obtained for $V_f^\text{JX}(r)$ (see Fig.~\ref{fig:res:QC1-normInt}), underlining the sensitivity of the total decay rates to the choice of the model potential.
The effective Coulomb potential $-1/r$ provides a qualitatively wrong description in both, the core and outer regions, whereas the $-6/r$ potential leads to a sort of compromise in accuracy.
It describes the core region much better than the $-1/r$ potential but as a trade off has a wrong asymptotic behavior in the valence region.
%
%
%
\bt{
For all but the $r^{-1}\sep \text{NO}$ approaches this suffices to predict spectra that are in qualitative agreement with the ones obtained for $V_f^\text{JX}(r)$, whereas using the $-1/r$ potential is only justified for the main features of the $d$ region, see \supp~Figs.~S2-S7.
}
%
%
If one is interested in the full spectrum, one should use a model taking into account the electronic potential of the ionized core, such as $V^{\rm scr}(r)$, $V_f^\text{J}(r)$ or $V_f^\text{JX}(r)$.
%
%

%
Summarizing the discussion until this point, it seems that satisfactory total decay rates and \acp{AES} are only obtained with the potentials $V_f^\text{J}(r)$ and $V_f^\text{JX}(r)$.
Further, employing $r^{-1}$ coupling and the \ac{SO} approximation seems to be very well justified for this models and we will use these in the comparison against experimental data below.
%
\bt{
Generally, special caution has to be taken with the inclusion of the NO overlap terms for simple potential models.
While the results obtained using $\mathcal{H}$ coupling greatly profit from error cancellation, this is not the case for $r^{-1}$ coupling, where the NO-terms strongly emphasize the deficiencies of an approximate potential.
In addition, it remains to be clarified, whether this cancellation effects are a general feature, or a peculiarity of the neon $1s^{-1}3p$ \ac{AES}.
}
%
%

%
%
\subsection{Comparison to experimental data}
\label{sec_res_disc:exp}
%
\begin{figure}

  \begin{minipage}{\columnwidth}
      \includegraphics{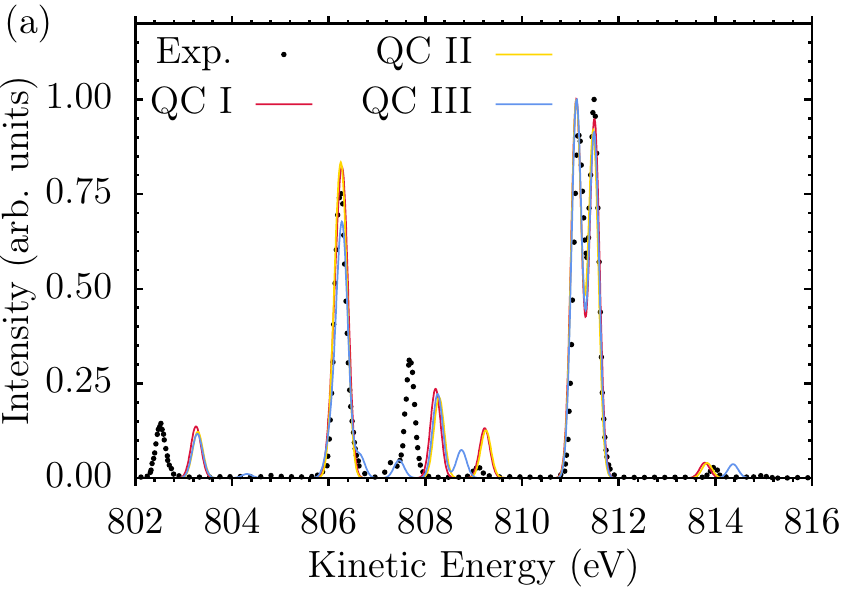}
  \end{minipage}
  \begin{minipage}{\columnwidth}
     \includegraphics{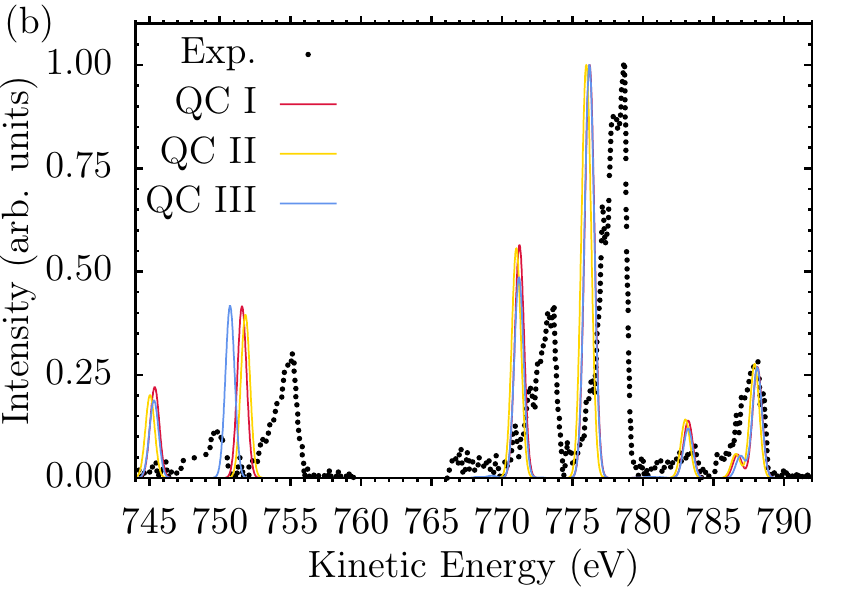}
  \end{minipage}
  \caption{\label{fig:res:exp} Comparison of experimental and theoretical neon $1s^{-1}3p$ \ac{AES} obtained based on the \ac{QC} schemes I - III with the $r^{-1} \sep V^\text{JX}_f(r) \sep \text{SO}$ method.
  The spectra obtained with \ac{QC} I, II, and III have been shifted by $-5.35$~eV, $-4.75$~eV, and $-5.18$~eV, respectively, to align the peak at $811.5$~eV with the experimental data in panel (a).
  %
  %
  To account for the different lineshapes of the experimental spectra that have been digitalized from \cite{Kivimaki_JESRP_2001}, panel (a), and \cite{Yoshida_JPBAMOP_2000} in panel (b), broadening with a Gaussian FWHM of $0.25$~eV and $ 0.77$~eV was used in panels (a) and (b), respectively.
  Further, the spectra have been normalized individually to the peaks at $811.5$~eV (a) and $776$~eV (b).
        }
\end{figure}
%
%
%
In this section, the comparison of our theoretical results to the experimental data in the full spectral range is presented.
In addition, to unravel the influence of the underlying \ac{QC} onto the \ac{AES}, we discuss spectra obtained with the \ac{QC} schemes I-III as described in Table~\ref{tab:QC}.
The \ac{QC} model II, which is more sophisticated than QC I, contains an uncontracted $s$ basis and accounts for scalar relativistic effects, whereas III employs an active space larger than in QC I .

To the best of our knowledge, no experimental Auger emission spectrum of the neon $1s^{-1}3p$ resonance that covers the full spectral range presented in Figs.~\ref{fig:res:QC1-normInt} and \ref{fig:res:QC1-waves} has been published to date.
Hence, in Fig.~\ref{fig:res:exp} the spectra obtained with the \ac{QC} models I-III and the $r^{-1} \sep V_f^\text{JX}(r) \sep \text{SO}$ approach are compared against experimental data taken from Kivim\"aki et al. \cite{Kivimaki_JESRP_2001}, for the $d$ region (a),
and Yoshida et al. \cite{Yoshida_JPBAMOP_2000}, for the $s$ and $p$ regions (b).
\bt{Note that the $\mathcal{H} \sep V_f^\text{JX} \sep \text{NO}$ approach does not lead to a considerable improvement of the agreement with the experimental data, as shown in \supp~Fig.~S9.}
%
%
%
The spectra have been shifted by $-5.35$~eV, $-4.75$~eV,  and $-5.18$~eV for \ac{QC} models I, II, and III to align the peaks at $811.5$~eV.
In panel (a), the spectra have been broadened using a Gaussian profile with an FWHM of $0.25$~eV, corresponding to the lineshape of the peak at $806.5$~eV in the experimental spectrum.
%
Further, in panel (b) a Gaussian FWHM of $0.77$~eV \bt{was chosen to represent the shape of the high energy flank of the asymmetric peak at $778.5$~eV in the experimental spectrum}.
Additionally, the spectra have been normalized to the heights of the main peaks at $811.5$~eV (a) and $776$~eV (b).
Finally, the experimental data have been recorded at angles of $54.7^\circ$ (a) and $56^\circ$ (b) to the polarization vector.
Data for these angles has been chosen to rule out anisotropy effects which makes the comparison with our angle integrated spectra easier.
Panel (a) shows that the agreement between the experimental and the theoretical spectra is fairly good.
In fact, the relative energetic positions and intensities of the main peaks are reproduced quite well.
However, smaller satellite peaks at $803$~eV and $808.2$~eV are blue shifted by about $0.5$~eV.
Generally, the agreement between theory and experiment is worse for the smaller features, although an unambiguous assignment is still possible.
Using the uncontracted $s$ basis and including scalar relativistic effects in \ac{QC} II does not influence the resulting spectrum.
In turn, the larger active space incorporated in the \ac{QC} scheme III leads to a better reproduction of some tiny features at $806.4$~eV, $807.5$~eV and $808.8$~eV.
The blue shifts of the peaks already present with \ac{QC} I and II are not affected, when employing \ac{QC} III.
The comparison of the spectra covering the $s$ and $p$ regions with the experimental data is shown in panel (b).
Here, the overall agreement is worse than in panel (a) both concerning the relative intensities and energetic positions of the peaks.
\bt{Regarding the offset in energies, the low-energy part corresponds to transitions to the highest excited states of the ionic manifold, the energies of which are progressively overestimated.
This is a typical situation when a smaller part of the electron correlation is recovered for the higher-lying states.}
Specifically, the positions of the peaks around $783$~eV and $788$~eV are reproduced well by all methods, \bt{while the intensity of the former, small peak is overestimated}.
Further, the peaks at $771$~eV and $776$~eV are red shifted by about $2$~eV, and the relative intensity of the former is overestimated by approximately $30\%$ (\ac{QC} I and II) and \bt{$15\%$} (\ac{QC} III).
Finally, the peaks at $745$~eV and around $751$~eV, corresponding to the $s$ region are redshifted by $5$~eV (\ac{QC} I-III) and $4$~eV (\ac{QC} I, II), or $5$~eV (\ac{QC} III), respectively.
The intensities of these peaks are overestimated by about $30\%$ with respect to the experimental data.
%
%
%
\bt{Hence, in the $s$ and $p$ regions, our results agree better with the experimental reference than with the \ac{MCDF} spectrum.}
%
%
The total decay rates, however, are not visibly altered by the choice of either of the \ac{QC} schemes I, II or III (cf. \supp~Fig. S8).
To wrap up this discussion, we conclude that the \ac{RASSCF}/\ac{RASPT2} electronic structure method combined with the $r^{-1} \sep V_f^\text{JX} \sep \text{SO}$ approach to construct continuum orbitals and evaluate the transition matrix elements provides Auger energies and intensities for the decay of the neon $1s^{-1}3p$ resonance of a similar quality as those obtained in Ref.~\citenum{Stock_PRA_2017} with the \ac{MCDF} approach.
In particular, a straightforward assignment of experimental results is possible.
Further, the inclusion of scalar relativistic effects into the one-component electronic structure of the bound states has no notable influence on the spectra, while a large active space is necessary only to reproduce minor satellite features.
%
\section{Conclusions and Outlook}
\label{sec_conclusion}
%
In this work, we have demonstrated an approach to the evaluation of autoionization rates on the example of the Auger decay from the neon $1s^{-1}3p$ resonance.
The suggested protocol is based on the \ac{RASSCF}/\ac{RASPT2} method to evaluate the bound state wave functions and energies, supplemented by a single-channel scattering model for the outgoing electron.
Here, the single-center approximation is introduced to reduce the continuum orbital problem to the radial dimension by averaging over the angular structure of the ionized electron density.
To model the true radial potential, six different models, $V^\text{free}$, $-1/r$, $-6/r$, $V^\text{scr}_f(r)$, $V^\text{J}_f(r)$, and $V^\text{JX}_f(r)$, have been discussed.
Further, three different ways to account for the nonorthogonality of the continuum and bound orbitals, \ac{SO}, NO, and \ac{GS}, as well as the effect of using complete, $\mathcal{H}$, or approximate, $r^{-1}$, coupling in the partial rate evaluation has been investigated.
All combinations of these sum up to 36 different variants to evaluate partial autoionization rates for an underlying bound state \ac{QC} calculation and have been implemented in a standalone program.
Here we compared all these approaches with respect to their ability to reproduce the experimental~\cite{Kivimaki_JESRP_2001, Yoshida_JPBAMOP_2000} as well as theoretical \ac{AES} obtained at the fully relativistic \ac{MCDF} level~\cite{Stock_PRA_2017}.
%
\bt{The applied quantum chemistry protocol allows for a fairly good reproduction of the transition energies if compared to the theoretical reference and experiments.
%
However, intensities are more difficult to reproduce.
Here,} the quality of the continuum orbital was shown to be the most important issue as it strongly influences the obtained \acp{AES}.
\bt{Especially the core part of the ionic potential is of importance for \ac{AES}.}
\bt{The effect of the potential is closely connected to the angular momentum of the outgoing electron as it governs the extent of the continuum orbital into the core region.}
For instance, we found that the $d$ region of the spectrum is rather insensitive to the choice of the model potential, whereas the $s$ and $p$ regions require to use one of the potentials $V^\text{scr}_f(r)$, $V_f^\text{J}(r)$ and $V^\text{JX}_f(r)$.
Still, the \ac{MCDF} intensities can only be reproduced in the $d$ region of the spectrum, while they are overestimated in the $s$ and $p$ parts.
Further, the free particle model and the asymptotic Coulomb potential $-1/r$ fail to reproduce the complete spectrum.
%
%
%
%
%
%
\bt{
Interestingly, inclusion of the NO terms in addition to using $r^{-1}$ coupling in the \ac{SO} approximation does not in general lead to improved spectra, but rather strongly emphasizes the deficiencies of the $V^\text{free}$, $-1/r$ and $-6/r$ potentials.
Due to pronounced error cancellation, however, this is diminished to large extent if the $\mathcal{H}$ coupling is used with the NO terms.
Remarkably, the effective $-6/r$ potential yields qualitative agreement with the spectra obtained using the more accurate potentials for all approaches but the aforementioned combination of $r^{-1}$ coupling and NO terms.
}
%
%
In contrast, the spectra obtained with $V^\text{scr}_f(r)$, $V_f^\text{J}(r)$ and $V^\text{JX}_f(r)$, are weakly affected by the choice of both the coupling and nonorthogonality treatment.
%
%
%
\bt{
Since it remains unclear whether the cancellation effects observed when using $\mathcal{H}$ coupling with the NO terms are a general feature, or a peculiarity of the neon \ac{AES}, the NO terms should only be employed with caution.
Generally, they should only be used, if the description of the potential and radial waves are sufficiently accurate.
When dealing with approximate potentials however, our results suggest to employ the \ac{SO} or \ac{GS} approaches, that seem to be less sensitive to the quality of  the continuum orbital.
In addition, due to the computational simplicity, the \ac{SO} approximation could be preferred.
}
The comparison with experimentally obtained spectra using the $r^{-1} \sep V_f^\text{JX}(r) \sep \text{SO}$ approach demonstrated the ability of the present method to accurately predict the neon $1s^{-1}3p$ \ac{AES}, allowing a straightforward assignment of the experimental data.
Interestingly, the general structure of the spectrum can be already reproduced quite well using a rather small active space, and is not sensitive to the inclusion of scalar relativistic effects.
%
%
The best agreement with the experimental data is achieved by using a larger active space, including additional excitations to $3d$, $5s$, $5p$, $6s$, and $6p$ orbitals together with the spherically averaged direct exchange potential $V_f^\text{JX}(r)$.
%
%
\bt{In addition our approach not only reproduces the experimentally measured \acp{AES} but the total decay rates of the neon $1s^{-1}3p$ resonance as well, when the potentials $-6/r$, $V_f^\text{J}(r)$, and $V^\text{JX}_f(r)$ are used.}
Since using the screened charge potential $V_f^\text{scr}(r)$ leads to notably underestimated absolute rates and is not computationally cheaper than using either $V_f^\text{J}(r)$ or $V^\text{JX}_f(r)$, providing a better accuracy, the latter two should be preferred. 
%
%
%
%

%
%
Molecular systems can be treated with the presented method as well,
\bt{however, in this case the molecular continuum has to be approximated by a single-centered spherically symmetric model.}
%
%
To keep the errors due to this approximation 
as small as possible, our findings suggest to use the  $r^{-1}$ coupling together with the \ac{SO} approximation to evaluate molecular \ac{AIS}.
The potentials should be modeled using either the direct $V_f^\text{J}(r)$ or direct-exchange $V_f^\text{JX}(r)$ variant.
The applicability of these approximations 
has to be tested for the molecular case.
Currently our code is interfaced to the MOLCAS/openMolcas as well as to the Gaussian program packages, allowing to evaluate \ac{PES} and \ac{AIS} based on bound state calculations conducted with the \ac{RASSCF}/\ac{RASPT2} \cite{Grell_JCP_2015}, as well as the linear-response time-dependent density functional theory method \cite{Mohle_JCTC_2018}.
A publication discussing the applicability of the present models to treat molecular systems will follow.
%
%
%
\appendix
\section{Obtaining the model potentials}
\label{sec:Ap_pot}
%
\bt{
For the state-dependent models, Eqs.~\eqref{eq:pot:scr}-\eqref{eq:pot:dirX}, the central quantity is the spherically-averaged electron density of the ionized state
\begin{equation}
 \label{eq:avgDens}
 \rho_f(r) = \frac{1}{4\pi}\int_0^{4\pi}\rho_f(\mathbf{r})\, d\Omega\,. 
\end{equation}
It determines the potentials in the following way:
The screened charge $Z_f(r)$ in Eq.~\eqref{eq:pot:scr} is evaluated as the difference between the nuclear charge $Z$ and the integrated number of electrons present in a sphere with radius $r$ around the atom:
\begin{equation}
\label{eq:screening}
Z_f(r) = Z - \int_0^r \rho_f(r')r'^2 dr'\,.
\end{equation}
%
Further, the direct Coulomb potential of the ionized states in  Eqs.~\eqref{eq:pot:dir} and \eqref{eq:pot:dirX} is
the electrostatic potential of the spherically averaged electron density $\rho_f(r)$,
\begin{equation}
  \label{eq:potJ}
  J_f(r) = J_f(0) - \frac{4\pi}{r}\int_0^r dr'\int_0^{r'}{r''\rho_f(r'') dr'' }\,.
\end{equation}
$J_f(0)$ is defined by the asymptotic value of the integral over $r''$ in Eq.~\eqref{eq:potJ}:
\begin{equation}
  \label{eq:J0}
  \frac{J_f(0)}{4\pi} = \lim_{r'\rightarrow\infty} \int_0^{r'}{r''\rho_f(r'') dr''}\,.
\end{equation}
Finally, a radial Slater type exchange \cite{Slater_PR_1951}, %
\begin{equation}
  \label{eq:potX}
  X_f^\text{S}(r)=-3\left(\frac{3}{8\pi}\rho_f(r)\right)^{\frac{1}{3}}\,,
\end{equation}
is employed.
}
\section{Coupling matrix elements}
\label{sec:Ap_TME}
%
\bt{
The total Auger transition matrix element $A_{i\alpha}^{\sigma M^+} = \mel{\Upsilon_\alpha^{\sigma M^+}}{\mathcal{H}-\mathcal{E}_i}{\Psi_i}$ reads
\begin{widetext}
  \begin{equation}
    A_{i\alpha}^{\sigma M^+} =
    \underbrace{
    \underbrace{\mel{\psi_{\alpha,\sigma}}{h}{\Phi_{i\alpha}^{M^+}}
                + \overbrace{
                  \sum_{q=1}^{N_{\rm orb}} \mel{\psi_{\alpha,\sigma}\varphi^+_{q}}{\frac{1}{r_{12}}}{\Xi_{i\alpha}^{M^+,q}}
                            }^{r^{-1}\text{ coupling}}
               }_{\text{SO}}
               + \braket{\psi_{\alpha,\sigma}}{\tilde{\Phi}_{i\alpha}^{1, M^+}}
               + \overbrace{
                            \braket{\psi_{\alpha,\sigma}}{\tilde{\Phi}_{i\alpha}^{2, M^+}}
                            }^{r^{-1}\text{ coupling}}
               - \mathcal{E}_i \braket{\psi_{\alpha,\sigma}}{\Phi_{i\alpha}^{M^+}}
               }_{\text{NO}}.
  \end{equation}
\end{widetext}
Here the contributions corresponding to the \ac{SO} approximation, the NO terms and the $r^{-1}$ coupling have been indicated.
If the GS approach is used, $\ket{\psi_{\alpha,\sigma}}$ is replaced by
\begin{equation}
  \label{eq:GS}
  \ket{\psi^\text{GS}_{\alpha,\sigma}} = \ket{\psi_{\alpha,\sigma}} - \sum_{i=1}^{N_\text{orb}}\braket{\varphi^+_{i,\sigma_i}}{\psi_{\alpha,\sigma}}\ket{\varphi^+_{i,\sigma_i}} .
\end{equation}
}
\bt{
The conjugated one and two-electron Dyson orbitals are defined as:
\begin{align}
  \label{eq:con1DO}
  \ket{\tilde{\Phi}^{1,M^+}_{i\alpha}} = \sum^{N_\text{orb}}_{q,s,t} \mel{\varphi^+_q}{h}{\varphi_s} \mel{\Psi^+_{f,M^+}}{(a^+_{q})^\dag a_s a_t}{\Psi_i} \ket{\varphi_t},
\end{align}
and
\begin{align}
  \label{eq:con2DO}
  \ket{\tilde{\Phi}^{2,M^+}_{i\alpha}} & = \sum^{N_\text{orb}}_{q_1<q_2}\sum^{N_\text{orb}}_{s_1<s_2,t} \Biggl(\mel{\varphi^+_{q_1} \varphi^+_{q_2}}{\frac{1}{r_{12}}}{\varphi_{s_1}\varphi_{s_2}} \\
  &- \mel{\varphi^+_{q_1} \varphi^+_{q_2}}{\frac{1}{r_{12}}}{\varphi_{s_2}\varphi_{s_1}}\Biggr) \nonumber\\
  &\times \mel{\Psi^+_{f,M^+}}{(a^+_{q_1})^\dag (a^+_{q_2})^\dag a_{s_2} a_{s_1} a_t}{\Psi_i} \ket{\varphi_t}. \nonumber
\end{align}
Note that the sum of both terms, which occurs when the $\mathcal{H}$ coupling is used, takes the simple form
\begin{equation}
  \label{eq:sumConDO}
  \ket{\tilde{\Phi}^{1,M^+}_{i\alpha}} + \ket{\tilde{\Phi}^{2,M^+}_{i\alpha}} = \mathcal{E}_f \ket{\Phi^{M^+}_{i\alpha}},
\end{equation}
as demonstrated in \cite{Manne_CP_1985}, if the Hamiltonian eigenvalue equation $\mathcal{H}\ket{\Psi^+_{f,M^+}} = \mathcal{E}_f\ket{\Psi^+_{f,M^+}}$ is used.
However, as detailed in the \supp~Section II, both approaches are strictly equivalent only, if the occupied spin-orbitals of the bound ionized and unionized wave functions span the same space.
This is generally also not fulfilled for \ac{RASSCF} orbitals if separate \ac{SCF} procedures are used to obtain ionized and unionized states since the respective transformation properties between orbitals from different subspaces are lost.
Moreover, the formulation of Ref.~\citenum{Manne_CP_1985} cannot be applied to the $r^{-1}$ coupling case whereas our formulation can.}
%
\begin{acknowledgments}
  We would like to thank Prof. Dr. Stephan Fritzsche, from the Helmholtz Institute Jena, Randolf Beerwerth and Sebastian Stock from his group for fruitful discussions and for providing the unshifted results of the MCDF calculations from Ref.~\citenum{Stock_PRA_2017}.
  Financial support from the Deutsche Forschungsgemeinschaft Grant No. BO 4915/1-1 is gratefully acknowledged.
\end{acknowledgments}


\newpage

\bibliography{neon_autoionization-rev-1.bib}

\end{document}